\documentclass[useAMS,usenatbib,a4]{mn2e}

\usepackage{times}
\usepackage{graphicx,bm,amssymb}
\usepackage{ulem}
\usepackage{epsfig}
\usepackage{amssymb}
\usepackage{natbib}
\usepackage{pstricks}
\usepackage{psfrag}
\usepackage[colorlinks=true,linkcolor=blue,citecolor=blue]{hyperref}
\def\apj{ApJ}
\def\apjl{ApJ}
\def\apjs{ApJS}
\def\aap{A\&A}
\def\mnras{MNRAS}
\def\pasa{PASA}
\def\prc{Phys.~Rev.~C}
\def\prd{Phys.~Rev.~D}
\def\rmxaa{Rev. Mexicana Astron. Astrofis.}
\def\physrep{Phys.~Rep.}

\newcommand{\be}{\begin{equation}}
\newcommand{\ee}{\end{equation}}
\newcommand{\bary}{\begin{eqnarray}}
\newcommand{\eary}{\end{eqnarray}}

\def\bi{\begin{itemize}}
\def\ei{\end{itemize}}
\def\lsim{\mathrel{\rlap{\lower3pt\hbox{\hskip1pt$\sim$}}
     \raise1pt\hbox{$<$}}} 
\def\gsim{\mathrel{\rlap{\lower3pt\hbox{\hskip1pt$\sim$}}
     \raise1pt\hbox{$>$}}} 

\begin{document}
\title[A Central Compact Object in Kes 79: The hypercritical regime and neutrino expectation]
{A Central Compact Object in Kes 79: The hypercritical regime and neutrino expectation}
\author[C. G. Bernal and N. Fraija]
  {C.~G. Bernal $^1$\thanks{E-mail:cgbernal@furg.br} and N.~Fraija$^2$\thanks{E-mail:nifraija@astro.unam.mx} \\
$^1$ Instituto de Matem\'atica, Estat\'istica e F\'isica, Universidade Federal do Rio Grande, Av. It\'alia km 8 Bairro Carreiros, Rio Grande RS, Brazil \\   
$^2$Instituto de Astronom\' ia, Universidad Nacional Aut\'onoma de M\'exico, Circuito Exterior,
C.U., A. Postal 70-264, 04510 M\'exico D.F., M\'exico}
                
\maketitle
\date{\today} 	
\begin{abstract}
We present magnetohydrodynamical simulations of a strong accretion onto magnetized proto-neutron stars for the Kesteven 79 (Kes 79) scenario. The supernova remnant Kes 79, observed with the Chandra ACIS-I instrument during approximately 8.3 h, is located in the constellation Aquila at a distance of 7.1 kpc in the galactic plane. It is a galactic and a very young object with an estimate age of 6 kyr. The Chandra image has revealed, for the first time, a point-like source at the center of the remnant. The Kes 79 compact remnant belongs to a special class of objects, the so-called Central Compact Objects, which exhibits no evidence for a surrounding pulsar wind nebula. In this work we show that the submergence of the magnetic field during the hypercritical phase can explain such behavior for Kes 79 and others CCOs. The simulations of such regime were carried out with the adaptive-mesh-refinement code FLASH in two spatial dimensions, including radiative loss by neutrinos and an adequate equation of state for such regime. From the simulations, we estimate that the number of thermal neutrinos expected on the Hyper-Kamiokande Experiment  is 733$\pm$364. In addition, we compute the flavor ratio on Earth for a progenitor model. 
\end{abstract}
\section{Introduction}\label{sec-Intro}
%
Central Compact Objects (CCOs) are point-like sources located at central regions of some supernova remnants, which have been observed only  in the X-rays \citep{2004IAUS..218..239P}. The X-ray spectra have been characterized to have  thermal components with blackbody temperatures of 0.2 - 0.5 keV and luminosities $L_X\sim 10^{33} - 10^{34}\,  {\rm erg\, s^{-1}}$. Several of such sources  and their supernova remnants (SNRs) are well-known, including RCW103, Cassiopeia A (Cas A), Pup A, and Kes 79 \citep{2010PNAS..107.7147K}. Currently, it is widely accepted that CCOs are neutron stars born in supernova explosions with an unusually estimated low magnetic field. A possible explanation for this atypical behavior is the so-called hidden magnetic field scenario. It suggests that the strong magnetic field has been buried due to a strong accretion during a core-collapse supernova (see e.g., \cite{1995ApJ...440L..77M}, \cite{1999A26A...345..847G}, \cite{Bernal2013} and references therein). When the core-collapse supernova event takes place, the shock is still pushing its way through the outer layers of the progenitor, and if it encounters a density discontinuity, a reverse shock may be generated. Depending on its strength and how far it was generated, this reverse shock can induce a strong accretion onto the newborn neutron star on a timescales of hours. Hypercritical accretion results when the accretion rate is higher than Eddington accretion rate ($\dot{M}>\dot{M}_{Edd}$). In this scenario, photons are trapped within the accretion flow and the energy liberated by the accretion is lost through neutrino emission close to the neutron star surface. After the reverse shock hits the neutron star surface and rebounds, a third shock develops and starts moving outward against the in-falling matter. Once this accretion shock stabilizes it will separate the in-falling matter from an extended envelope in quasi-hydrostatic equilibrium.
Based on such scenario of late accretion onto newborn neutron stars inside supernovae, \cite{1989ApJ...341..867C} developed an analytical model for the hypercritical regime. In this model, the neutrino cooling plays an important role in the formation of a quasi-hydrostatic envelope around the compact remnant.\\
In \cite{1989ApJ...346..847C}, the author highlighted some conditions involved in the formation of neutron stars inside supernovae, including the hyperaccretion of material due reverse shock and the formation of a convective envelope around the compact remnant. Following such ideas, \cite{1995ApJ...440L..77M} considered the submergence of the magnetic field on the stellar crust and the subsequent ohmic diffusion of the submerged magnetic flux through the outermost nonmagnetic layers of the crust of a newborn neutron star.
With these requirements, \cite{1999A26A...345..847G} presented simple 1D ideal magnetohydrodynamical (MHD) simulations of the effect of this post-supernova hypercritical accretion on the newborn neutron star magnetic field to show that such submergence/re-emergence could occur.  These simulations displayed a rapid submergence of the magnetic field into the neutron star.  Based on MHD simulations with high refinement,  \cite{Bernal2010,Bernal2013} confirmed this result.  After hyperaccretion stops, the magnetic field could diffuse back to the surface, resulting in a delayed switch-on of a pulsar. It is worth noting that depending on the amount of accreted material, the submergence could be so deep that the neutron star may appear and remain unmagnetized for centuries \citep{1999A26A...345..847G}.  Recently, some authors revisiting this scenario studied the magnetic field evolution in the CCOs context \citep{2011MNRAS.414.2567H, 2012MNRAS.425.2487V}.   In addition, \cite{Shabaltas2012} and \cite{Popov2015} have suggested that such hidden magnetic field scenario may be applicable for various CCOs, including the compact remnant in Kes 79. \\
Located in the Galactic plane at 33 degrees northeast of the Galactic center, Kes 79 (also known as G33.6+0.1) is the source 79 in the radio catalog of Kesteven \citep{Kesteven1968}.  It is a moderately large SNR with a point-like source at its center,  widely believed to be a neutron star created in the SN explosion. Such CCO  is called CXO J1852.6+0040 \citep{Seward2003}. The authors showed that the luminosity, in the X-ray band (0.3-8 keV), is $\sim 7\times10^{33}$ erg s$^{-1}$, which corresponds to four times the X-ray luminosity reported in the CCO  of Cas A. The blackbody spectrum peaking at X-rays and the lack of a Pulsar Wind Nebula (PWN) indicate that the thermal emission could be originated from a small region, perhaps on the surface of the neutron star. As \cite{Popov2015} pointed out, the hypothesis of their suppressed magnetosphere is mainly based on the analysis of their thermal emission: pulse profiles of the X-ray light curves and a high pulse fraction, which requires magnetar-scale fields in the crust.\\
\cite{2005ApJ...627..390G} reported the discovery of 105 ms X-ray pulsations from the CCO in Kes 79, with an upper limit on its spin-down rate of $\dot P<7 \times 10^{-14}$ s s$^{-1}$. Assuming a magnetic dipole model they estimate that the surface magnetic field strength is $B<3 \times 10^{12}$ G. Also, if a blackbody model of temperature $T_{BB}=(0.44 \pm 0.03$) keV is used for the X-ray spectrum characterization, they estimated a radius for the source of $R_{BB} \sim 0.9$ km. More recently, \cite{2014ApJ...790...94B} modeled such X-ray pulsations in Kes 79 in the context of thermal surface radiation from a rotating neutron star and more accurate results for the surface magnetic field were estimated.  Taking into account  the reasons mentioned above and the similarities between the CCO in Kes 79 and the neutron star in Cas A, in the present work we adopt, for simplicity, standard parameters for the neutron star in Kes 79: $M \simeq 1.44$ $M_{\odot}$, $R \simeq10$ km, and an average pre-hyperaccretion magnetic field of $B \simeq 10^{12}$ G. Such parameters are adequate for progenitor models of pre-supernova in the range 20--40 $M_{\odot}$ as it seems to be the case of Kes 79 (see \cite{Chevalier2005, Shabaltas2012} and references therein.) Although we have chosen the Kes 79 CCO as a particular case, the method presented here can be adapted for other CCOs with similar features.
\\
In the hidden magnetic field scenario, the neutrino production and cooling on the neutron star surface play an important role in the formation of a quasi-hydrostatic envelope \citep{Bernal2013, 2014MNRAS.442..239F, 2015MNRAS.451..455F}. The  properties of these neutrinos get modified when they propagate in this thermal and magnetized medium. For instance, depending on the flavor composition, neutrinos would feel a different effective potential. These changes in the flavor mixing  are due to the electron neutrino ($\nu_e$) interaction with electrons via both neutral and charged currents (CC), whereas muon ($\nu_\mu$) and tau ($\nu_\tau)$ neutrinos interact only via the neutral current (NC).  The resonant conversion of neutrino from one flavor to another caused by the medium effect is well-known as the Mikheyev-Smirnov-Wolfenstein effect \citep{wol78}.\\
%
%
Using the FLASH code to simulate the dynamics close to the stellar surface (including the reverse shock, the neutrino cooling processes and the magnetic field dynamics), some authors presented numerical studies of hypercritical accretion of matter onto the neutron star surface in the center of the SNRs 1987A and Cas A \citep{2014MNRAS.442..239F,2015MNRAS.451..455F}. The authors estimated the number of neutrinos that must have been  seen from the hypercritical accretion episode on the Super-Kamiokande (SK) and Hyper-Kamiokande (HK) neutrino experiments. In this work, we study the dynamics of the envelope and analyze the submergence of the magnetic field into the stellar surface for the compact remnant in Kes 79. We estimate the neutrino flux and the flavor ratio expected on Earth. The paper is arranged as follows. In section \ref{sec-Physics} we describe the physics included in the hypercritical model. In section \ref{sec-FLASH} we show the numerical results from MHD simulations. In section \ref{sec-Neutrinos} we study the thermal neutrino dynamics and we analyze the neutrino oscillations. Finally, a brief conclusions are drawn in section \ref{sec-Results}. 
%
%
\section{Physics input and numerical method}\label{sec-Physics}
 
\cite{Chevalier2005} showed that progenitors of Type-II supernova explosions are red supergiants which already have lost their hydrogen envelope via a powerful stellar wind. Then, it is inferred that such progenitors with masses in the range 20--40 $M_{\odot}$ must have a helium core of $\sim 6-7 \:\mathrm{M_{\odot}}$ and an iron core of $\sim 1.5-1.6\:\mathrm{M_{\odot}}$. In this  scenario,  a neutron star with $\simeq 1.4\:\mathrm{M_{\odot}}$ is created.  It seems to be the case of Kes 79. 
As aforementioned, the CCO in Kes 79 looks similar to that CCO present in Cas A \citep{Chakrabarty2001}. If the accretion onto the newly born neutron star surface is hypercritical ($\dot M\geq10^{4}\: \dot M_{Edd}$), where the Eddington accretion rate is $\dot M_{Edd} \sim 10^{18}$ gr s$^{-1}$, then the ram pressure of the falling matter can exceed by several orders of magnitude the superficial magnetic pressure of the neutron star, submerging the magnetic field into the new stellar crust very fast \citep{Bernal2010, Bernal2013, 2014MNRAS.442..239F}. These authors showed that the magnetic field is frozen into the matter and, under certain conditions for the accretion rate and magnetic field, initial configurations of the magnetic field on the neutron star can be rapidly submerged beneath the accreted matter.  For instance,  \cite{2016MNRAS.456.3813T} showed that typical magnetic fields of standard neutron stars can be buried by modest accreted mass of the order that $0.01-0.001\:\mathrm{M_{\odot}}$ in the hypercritical phase.
This suggests that neutron stars produced by supernovae, in which post core-collapse accretion is hypercritical, are born with a weak superficial magnetic field due to the bulk of the magnetic field was buried into the neutron star crust.
\subsection{Structure of the regions}
Four dynamically different regions can be identify in the pre-supernova: (i) a thin crust formed during the hypercritical phase; (ii) an envelope in quasi-hydrostatic equilibrium; (iii) a free-fall region; and (iv) the external layers of the progenitor. In this context, analytical approaches 1D can be used to estimate some parameters in the system. However, numerical simulations with more degrees of freedom and more physical ingredients are necessary to study in detail the magnetohydrodynamic close to the neutron star surface.
In table \ref{table1} we summarize the densities and radii of each region and also give a brief description as follows.
\begin{table*}       
\begin{center}
\caption[]{\small\sf Description of the regions during the hypercritical accretion phase}\label{table1}
\begin{minipage}{157mm}
\begin{tabular}{|c|c|c|c|c|c|}
\hline \hline
\multicolumn{5}{c}{}\\
 {\small Zones}  & {\small Density} &  {\small radii} & &\\
                         &  (${\rm g~cm}^{-3})$ & (cm) &\\\hline

{\small I. New crust of ns surface}      &  {\small $\rho_{e}(r_{c}/r)^n$}   &    {\small $r_{ns}=10^{6}\,\,$;  $\,\,r_{c}\sim 200-400$}   &  \\
{\small ($ r\leq [r_{ns}+r_c]$)}                             & & &\\\hline

{\small II. Quasi-hydrostatic envelope}     & {\small $7.7 \times 10^2 \,\left(\frac{r_{s}}{r}\right)^{3}$}  & {\small  $r_{s}\simeq7.73\times10^{8}$}\\
{\small ($[r_{ns}+r_c]\leq r\leq r_s$)}                            & & &\\\hline

{\small III. Free-fall}    &  {\small $5.74\times 10^{-2}\,\biggl(\frac{r}{r_h} \biggr)^{-3/2}$}     &  {\small $r_h= 6.3\times 10^{10}$}     & \\
{\small ($r_s\leq r\leq r_h$)}                             & & &\\\hline

{\small IV. External layers}    & {\small$ 3.4\times10^{-5} \times\cases{
\left(\frac{R_\star}{r}\right)^{17/7}; & $r_h< r < r_b$,\cr
\left(\frac{R_{\star}}{r}\right)^{17/7}\frac{\left(r-R_{\star}\right)^{5}}{\left(r_{b}-R_{\star}\right)^{5}}; & $r>r_b$.\cr
}
$}     &   {\small $r_{b}=10^{12}\,\,$;  $\,\,R_{\star}\simeq3\times10^{12}$}     &\\
{\small ($r_h\leq r\leq r_b\,$ and $\, r>r_b$)}                             & & &\\
\hline\hline
\end{tabular}
\end{minipage}
\end{center}
\end{table*}
\subsubsection{Zone I: New crust on the stellar surface} When the hypercritical phase takes place, the piled material forms a new crust  with a strong magnetic field immersed within it. During this phase the magnetic field in the range $10^{11}\,{\rm G} \leq B \leq 10^{13}\,{\rm G}$ may be submerged and confined into the new crust at $r_{ns}  \leq r \leq (r_{ns}+r_{c})$.  Here, $r_{ns}=10^{6}$ cm is the neutron star radius and $r_{c}\sim 200-400$ m represent the height-scale of the new crust formed by the accreted matter in the hypercritical phase. Previous simulations performed in \cite{Bernal2010, Bernal2013} showed that this is the case for typical parameters of newly born neutron stars. The density and pressure in the new crust can be modeled as power laws  {\small $\rho=\rho_{e}(r_{c}/r)^n$} and  {\small $p=p_{e}(r_{c}/r)^m$}, respectively, where $\rho_{e}$ and $p_{e}$ are the values at the base of the envelope and $n$ and $m$ are the power indexes, which depend on the accretion rate.  Radiation and neutrinos confined in this region are thermalized to a few MeV with a neutrino emissivity given by {\small $\dot{\epsilon}_{\nu}=0.97\times10^{25}(T/\mathrm{MeV})^{9}\:\mathrm{erg\, s^{-1}\, cm^{-3}}$} \citep{Dicus1972}.  Due to the high dependence of the pair annihilation process with pressure  near the stellar surface {\small $p_{e^{\pm}}=\frac{11}{12} aT^4\,$}, it is possible to write the neutrino emissivity as  a function  of total pressure ($p$) as {\small $\dot{\epsilon}_{\nu} = 1.83 \times 10^{-34} p^{9/4}\: \mathrm{erg \, cm^{-3} \, s^{-1}}$} with $a$ the radiation constant. 
\subsubsection{Zone II: Quasi-hydrostatic envelope}
The reverse shock induces hypercritical accretion onto the newborn neutron star surface. The accreted material bounces off  the stellar surface forming a new expansive shock, which builds an envelope in quasi-hydrostatic equilibrium around the neutron star, with free falling material raining over it. The structure of such envelope can be described through density $\rho_{qhe}=\rho_{s}\left(\frac{r_{s}}{r}\right)^{3}$,  pressure $p_{qhe}=p_{s}\left(\frac{r_{s}}{r}\right)^{4}$ and  velocity $v_{qhe}=v_{s}\left(\frac{r_{s}}{r}\right)^{-1}$ where the subscript \textit{s} refers to the value of density ($\rho_{s}$), pressure ($p_{s}$) and velocity  ($v_{s}$) at the shock front and the shock radius $r_{s}\simeq7.73\times10^{8}\mathrm{cm}$ \citep{1989ApJ...341..867C}.  The pressure $p_{s}$ and density $\rho_{s}$ are determined by the strong shock condition $p_{s}=\frac{6}{7}\rho_{0}v_{0}^{2}$ and $\rho_{s} = 7 \rho_{0}$, respectively and $v_{s}$ by the mass conservation $v_{s} = -\frac{1}{7} v_{0}$.   Here $\rho_{0}$ and $v_{0}$ are the density and velocity just outside the shock, respectively. It is worth noting that the quasi-hydrostatic envelope solution described here corresponds to a $\gamma=4/3$ polytrope. Using the analytical formula for the neutrino emissivity  \citep{Dicus1972},  the radial location of the accretion shock that is controlled by the energy balance between the accretion power and the integrated neutrino losses is {\small $GM\dot{M}/R=\pi R^{3}\dot{\epsilon}_{\nu}$}.  Then, the high pressure near the neutron star surface which is given by {\small $p_{ns}\simeq1.86\times 10^{-12} \mathrm {dyn\, cm^{-2}} \dot{M}\, r_{s}^{3/2}$} allows the pair neutrino process to be the dominant mechanism in the neutrino cooling.  From the strong shock conditions, we get {\small $\rho_{qhe}(r)= 7.7 \times 10^2 \,\left(\frac{r_{s}}{r}\right)^{3}~{\rm g~cm}^{-3}$}, where the quasi-hydrostatic envelope radius lies in the range $(r_{ns}+r_{c}) \leq r \leq r_s$.
\subsubsection{Zone III: The free-fall region} In the free-fall region, material start falling  with the velocity {\small $v_{ff}(r)=\sqrt{\frac{2GM}{r}}$} and density profiles $\rho_{ff}(r)=\frac{\dot{M}}{4\pi r^{2}v(r)}$.  Requiring the typical values for such object, $M\sim1.4 ~{\rm M_\odot}$ and $\dot{M}\sim 10^{3} ~{\rm M_{\odot}}\, {\rm yr^{-1}}$,  the velocity is {\small $v_{ff}(r)=7.81\times10^{7}\,(r/r_h)^{-1/2}\:\mathrm{cm\, s^{-1}}$} and the density of material in free fall is  {\small $\rho_{ff}(r)=5.74\times 10^{-2}\,(r/r_h)^{-3/2}~{\rm g~cm}^{-3}$} where $r_s\leq r\leq r_h$ with  $r_h= 6.3\times 10^{10}$ cm.
\subsubsection{Zone IV: The external layers} By considering a typical profile for a pre-supernova, the density of the external layers can be described as: $\rho_0\,\left(\frac{R_\star}{r}\right)^{17/7}$ for $r_h< r < r_b$ and $\rho_0\,\left(\frac{R_{\star}}{r}\right)^{17/7}\frac{\left(r-R_{\star}\right)^{5}}{\left(r_{b}-R_{\star}\right)^{5}}$ for $r>r_b$ with $\rho_0=3.4\times10^{-5}\mathrm{g\,cm^{-3}}$,  $r_{b}=10^{12}$ cm and  $R_{\star}\simeq3\times10^{12}$ cm  \citep{1989ApJ...341..867C}.
\subsection{Numerical method}

To tackle numerically the hypercritical accretion problem, we use a customized version of the Eulerian numerical code FLASH \citep{Fryxell2000}. We use the Split Eight-Wave solver to solve the whole set of MHD equations. The eight-wave magnetohydrodynamic solver is based on a finite-volume, cell-centered method that was proposed by \cite{powell1999solution}. The solver uses directional splitting to evolve the magnetohydrodynamics equations. It makes one sweep in each spatial direction to advance physical variables from one time level to the next. In each sweep, the unit uses AMR functionality to fill in guard cells and impose boundary conditions. Then, it reconstructs characteristic variables and uses these variables to compute time-averaged interface fluxes of conserved quantities. In order to enforce conservation at jumps in refinement, the solver makes flux conservation calls to AMR, which redistributes affected fluxes using the appropriate geometric area factors. Finally, the solver updates the solution and calls the EOS unit to ensure thermodynamical consistency. A difficulty particularly associated with solving the MHD equations numerically lies in the solenoidality of the magnetic field. The notorious $\nabla\cdot\mathbf{B}=0$ condition, a strict physical law, is very hard to satisfy in discrete computations. Being only an initial condition of the MHD equations, it enters the equations indirectly and is not, therefore, guaranteed to be generally satisfied unless special algorithmic provisions are made. FLASH provides the truncation-error cleaning method for the eight-wave MHD unit as was described in \cite{powell1999solution}. We work in the ideal MHD regime with only numerical resistivity and viscosity. This is justified due to the violence and short duration of the hypercritical phase.
FLASH code solves the equations of compressible ideal and non-ideal magnetohydrodynamics in one, two and three dimensions on a Cartesian system. The code is suitable for simulations of ideal plasmas in which magnetic fields can be  so strong (or weak) that they do not cause temperature anisotropies (i.e. physically consistent). In principle, they can have any initial configuration.

Here, we are interested in analyzing the formation of the envelope around the neutron star and the subsequent building of the new crust (with the magnetic field confined inside), from the initial reverse shock in free-fall until the establishment of the atmosphere in quasi-hydrostatic equilibrium. The external layers are not dynamically important in this part of the work, and therefore not considered in the simulations.
\subsubsection{Equation of state}
The thermal pressure is a scalar quantity and as given in regular hydrodynamics, it is obtained from the internal energy and density using the equation of state.
The matter equation of state is an adaptation of FLASH's Helmholtz package which includes contributions from the nuclei, $e^\pm$ pairs, radiation and the Coulomb correction. This routine is appropriate for addressing astrophysical phenomena in which electrons and positrons may be relativistic and/or degenerate, and radiation may significantly contribute to the thermodynamic state. The Helmholtz Unit in FLASH provides a table of the Helmholtz free energy (hence the name) and makes use of a thermodynamically consistent interpolation scheme obviating the need to perform the complex calculations required of the intrinsic formalism during the course of a simulation. The interpolation scheme uses a bi-quintic Hermite interpolant resulting in an accurate EOS that performs reasonably well.  A detail description of the Helmholtz equation of state is provided by \cite{Timmes2000}.  The range of validity of the Helmholtz Unit in the FLASH code is

\begin{equation}
10^{-10}<{\rho}<10^{10}\,\, {\rm  g\, cm^{-3}} \hspace{0.5cm}{\rm and} \hspace{0.5cm}10^{4}<{T}<10^{10}\,\, {\rm  K}\,.
\end{equation}

\noindent As in our analysis we  exceed this range, we make an extension to the equation of state for degenerate regime ($\rho, T>10^{10}$ units).

\paragraph{Ions.} In this case (degenerated regime) we propose a simple analytical relation for the specific energy, pressure and entropy ($\varepsilon^{ions}_{dg}, P^{ions}_{dg}, S^{ions}_{dg}$) and their gradients, which fits successfully the numerical results of equations of state for degenerate matter as indicated in \cite{Timmes2000}:
\begin{eqnarray}
\varepsilon _{dg}^{ions} &=&k\rho \:\: \rightarrow \:\: \frac{%
d\varepsilon }{d\rho }=k,\:\: \frac{d\varepsilon }{dT}=0, \\
P_{dg}^{ions} &=&k\rho ^{2}\:\: \rightarrow \:\: \frac{dP}{d\rho }%
=2k\rho,\:\: \frac{dP}{dT}=0, \\
S_{dg}^{ions} &=&0\:\: \rightarrow \:\: \frac{dS}{d\rho }=0,\:\: \frac{dS}{dT}=0,
\end{eqnarray}
\noindent where $k=2.7941\times 10^{4}$ is a proportionality constant calculated in the transitional regime and the subindex {\it dg} refers to degeneracy.
\paragraph{Electron-positron pairs ($e^\pm$).}
The total specific electron energy and its gradients are calculated through the degenerated and pair contributions, which are given by
\begin{eqnarray}
\varepsilon _{elec} &=&\sqrt{\varepsilon _{dg}^{2}+\varepsilon _{par}^{2}},
\end{eqnarray}
and
\begin{eqnarray}
\frac{d\varepsilon _{elec}}{d\rho } &=&\frac{\varepsilon _{dg}\frac{%
d\varepsilon _{dg}}{d\rho }+\varepsilon _{par}\frac{d\varepsilon _{par}}{%
d\rho }}{\varepsilon _{ele}}, \\
\frac{d\varepsilon _{elec}}{dT} &=&\frac{\varepsilon _{dg}\frac{d\varepsilon
_{dg}}{dT}+\varepsilon _{par}\frac{d\varepsilon _{par}}{dT}}{\varepsilon
_{ele}},
\end{eqnarray}
%
%
where the contribution of degeneracy is given by {\small $\varepsilon _{dg} =\frac{3}{4}\frac{n_{e}}{\rho }E_{F} [ 1+\frac{2}{3} \pi ^{2} ( k_{B}T/E_{F})^{2}] =A\rho ^{1/3}(1+B\rho^{-2/3}T^{2})$}, where $A$ and $B$ are given by  {\small $A = \frac{3 Y_{e}^{4/3} ( 3\pi ^{2})^{1/3}}{4m_{H}^{4/3}} $},  {\small $B= \frac{3\pi k_{B}m_{H}^{1/3}}{(24\pi ^{2}Y_{e}) ^{1/3}}$},  {\small $n_{e}=Y_{e}\rho /m_{H}$}  is the electronic density, {\small $m_{H}$} being the hydrogen mass, {\small $Y_{e}=Z/A$} is the electronic fraction and {\small $E_{F}=(3\pi ^{2})^{1/3}n_{e}^{1/3}$} is the Fermi energy.  The contribution of pairs is {\small $\varepsilon _{par} =\frac{7}{4}\varepsilon _{rad}=\frac{7}{4}\frac{aT^{4}}{\rho }$}. Taking into account the previous relations, the gradients of the high degeneration contribution for the specific energy can be written as
\begin{eqnarray}
\frac{d\varepsilon _{dg}}{d\rho } &=&\frac{1}{3}A\rho ^{-2/3}\left( 1-B\rho
^{-2/3}T^{2}\right) , \\
\frac{d\varepsilon _{dg}}{dT} &=&2AB\rho ^{-1/3}T,
\end{eqnarray}
\noindent and the pairs contribution is
\begin{eqnarray}
\frac{d\varepsilon _{par}}{d\rho } &=&-\frac{7}{4}\frac{aT^{4}}{\rho ^{2}}=-%
\frac{\varepsilon _{par}}{\rho }, \\
\frac{d\varepsilon _{par}}{dT} &=&7\frac{aT^{3}}{\rho }=\frac{4}{T}%
\varepsilon _{par}.
\end{eqnarray}
\noindent  On the other hand, the pressure and its gradient associated is given by
\begin{eqnarray}
P_{dg} &=&\frac{1}{3}\rho \varepsilon _{ele}, \\
\frac{dP}{dT} &=&\frac{1}{3}\rho \frac{d\varepsilon _{ele}}{dT},\:\:\:%
\frac{dP}{d\rho }=\frac{1}{3}\varepsilon _{ele}.
\end{eqnarray}
Finally, the total entropy is
\begin{equation}
S_{T}=\sqrt{S_{dg}^{2}+S_{par}^{2}},
\end{equation}
and their gradients can be written as 
\begin{eqnarray}
\frac{dS_{T}}{d\rho } &=&\frac{S_{dg}\frac{dS_{dg}}{d\rho }+S_{par}\frac{%
dS_{par}}{d\rho }}{S_{T}}, \\
\frac{dS_{T}}{dT} &=&\frac{S_{dg}\frac{dS_{dg}}{dT}+S_{par}\frac{dS_{par}}{dT%
}}{S_{T}}.
\end{eqnarray}
\noindent where the entropy of  the degeneracy is {\small $S_{dg} =\int_{0}^{T}\frac{C_{V}}{T}dT=2AB\rho ^{-1/3}T=\frac{d\varepsilon _{dg}}{dT}$} with  the heat capacity at constant volume $C_{V}=\frac{d\varepsilon _{dg}}{dT}=2AB\rho ^{-1/3}T$ and the contribution of pairs to the entropy is {\small $S_{par}= \frac{7}{4T}\left[ \frac{P_{rad}}{\rho }+\varepsilon _{rad}\right] = \frac{7}{3}\left[ \frac{aT^{3}}{\rho }\right] $}. The gradients to the case of degeneracy and pairs are
\begin{eqnarray}
\frac{dS_{dg}}{d\rho } &=&-\frac{2}{3}AB\rho ^{-\frac{4}{3}}T=-\frac{S_{dg}}{%
3\rho }, \\
\frac{dS_{dg}}{dT} &=&2AB\rho ^{-\frac{1}{3}}=\frac{1}{T}S_{dg}\\
\frac{dS_{par}}{d\rho } &=&\frac{7}{4}\frac{dS_{rad}}{d\rho }=-\frac{S_{par}%
}{\rho }, \\
\frac{dS_{par}}{dT} &=&7\frac{aT^{2}}{\rho }=\frac{3}{T}S_{par}.
\end{eqnarray}

\noindent Implementing these corrections in the Helmholtz Unit of the FLASH code, we take into account the contributions of degenerate and relativistic particles for the equation of state, with the conditions $\rho >10^{10}$ g cm$^{-3}$ and $T>10^{10}$ K.\\
\subsubsection{Neutrino Cooling Processes}
Simulations of the neutrino emission and their interactions have found that they carry out a leading role in the dynamics of supernovae and the recently born neutron star so-called proto neutron star (pns) (see \cite{doi:10.1146/annurev-nucl-102711-095006, 2015PhRvC..91c5806R} and references therein).    In the hypercritical accretion phase the neutrino energy losses are dominated by the annihilation process ($e^{+}-e^{-}\rightarrow\nu+\bar{\nu}$),  which involves the formation of a neutrino-antineutrino pair. In the present work we include additionally other relevant neutrino processes present in such regime: (i) {\bf the photo-neutrino process}. This process is described by the outgoing photon in a Compton scattering going to a neutrino-antineutrino pair ($\gamma+e^{\pm}\rightarrow e^{\pm}+\nu+\bar{\nu}$); (ii) {\bf the plasmon decay process}.  It occurs when a photon propagating within an electron gas (plasmon) is spontaneously transformed in a neutrino-antineutrino pair ($\gamma\rightarrow\nu+\bar{\nu}$); and (iii) {\bf the Bremsstrahlung process}.  It takes place when a photon is replaced by a neutrino-antineutrino pair, either due to electron-nucleon interactions ($e^{\pm}+N\rightarrow e^{\pm}+N+\nu+\bar{\nu}$) or nucleon-nucleon interactions ($N+N\rightarrow N+N+\nu+\bar{\nu}$).    All these processes have been well described by \cite{Itoh1996}. In their work, the authors have summarized the results of the calculations of the neutrino energy-loss rates resulting from pair, photo-, plasma, bremsstrahlung, and recombination neutrino processes based on the Weinberg-Salam theory \citep{BILENKY198273}. A wide density-temperature regime $1\leq\rho/\mu_{e}\leq10^{14}$ g cm$^{-3}$ and $10^{7}\leq\ T \leq10^{11}$ K has been considered. In the present work, we use the extensive numerical tables for the neutrino processes above mentioned as FORTRAN routines, including the
interpolation formulae of the tables and the analytic fitting formulae, implemented in a customized module in FLASH. We consider initially radial profiles for the pressure and density, implying a radial distribution for the energy. However, in the neutrinosphere (i.e. the region where neutrinos are produced) we have assumed that mean neutrino energy is around $k_{B} T$ and all emitted neutrinos have this energy. The numerical routine apply the neutrino-cooling source term operator to a block of zones. The neutrino losses rate is used to update the internal energy in the zone. After we call neutrino-cooling routine, call the EOS to update the pressure and temperature based on the neutrino losses. This implementation in FLASH prevents stiffness problems that may arise in the simulation.
Moreover, the main sources of neutrino opacity are then coherent scattering of neutrons and protons, and pair annihilation. For example, the corresponding cross section for coherent scattering is $\sigma_{\rm N}=(1/4)\sigma_0[E_{\nu}/(m_e c^2)]^2$, where $\sigma_0=1.76 \times 10^{-44}$~cm$^{2}$. As neutrinos are created in a thermal medium with temperatures $T \lesssim 10^{11} {\rm K} \lesssim 8$ MeV, their energies can be estimated by $E_\nu \sim k_{B} T$ and the cross section would be $\sigma_{\rm N} \lesssim 7 \times 10^{-42}$ cm$^2$. The maximum densities reached at the bottom of the envelope becomes $\rho \sim 10^{10}$ g cm$^{-3}$, and in such conditions, the neutrino mean free-path would be $l_{\nu}=(n_{\rm N} \sigma_{\rm N})^{-1} \gtrsim 2.5 \times 10^{6}$~cm, which is safely larger than the depth of the dense envelope (of order of a few km). Above this dense region the envelope density decreases rapidly and the whole envelope is practically transparent to neutrinos. We will, hence, ignore neutrino absorption and heating.
As in previous works, no nuclear reactions are taken into account.

\subsubsection{Initial and boundary conditions}

As explained in \cite{Bernal2013}, we want to simulate a portion of the surface of the neutron star (spherical geometry) doing a mapping of the surrounding cone above the neutron star into an accreting column (cartesian geometry).  Due to the MHD equations it can be successfully solved by FLASH in cartesian coordinates and also the computational cost is minor. This implies that all variables and parameters of the physical model are mapped from a spherical to a cartesian geometry, without loss of generality.
For our 2-dimensional computational domain we consider wide columns with a base of $\Delta x = 2 \times 10^6$ cm (centered on $x=0$), and a height $\Delta y$ of $40 \times 10^6$ cm, with $y=10^6$ cm being the neutron star radius. 
We chose this height-scale to prevent the shock from leaving the computational domain during the initial transient. The gravitational acceleration (plane-parallel external field) is taken as $g_y=-GM/y^2$ and  a standard neutron star mass of $1.44\: M_\odot$ is assumed.\\ 
As magnetic initial condition, we consider a magnetic field loop, in the shape of an hemi-torus. The reason is that, as showed in \cite{Bernal2010, Bernal2013}, under certain conditions of accretion rate, any initial magnetic field configuration will be submerged in the hypercritical regime by the accreted matter. However, a magnetic loop is very interesting because the magnetic tension due to the magnetic field curvature can play an important role in the plasma-dynamics very close to the stellar surface. 
On the central hemi-circle of the loop the field has a strength $B_0 = 10^{12}$ G (standard for neutron stars) and about it is shaped as a Gaussian, i.e., with strength, 
\begin{equation}
B(d) = B_0\times \exp\left(-\frac{d}{R_L}\right)^2
\label{Eq:loop}
\end{equation}
where $d$ is the distance to the loop central hemi-circle and the thickness of the loop is $R_L = 1$ km. The two feet of the loop are centred at $x=-5$ and $x=+5$ km. As boundary conditions, we impose mass inflow along the top edge of the computational domain with constant hypercritical accretion rate, and periodic conditions along the sides, allowing fluid to move freely through it. 
At the bottom, on the neutron star surface, we use a custom boundary condition which enforces hydrostatic equilibrium. In order to establish this boundary we fix the velocities as null in all the guard cells, $(v_{x}=v_{y}=0),$ and copy the density and the pressure of the first zone of the numerical domain. This zone corresponds to the neutron star surface $\left( \rho =\rho(ic), p=p(ic)+\rho v^{2}+\rho gy\right)$, where $ic$ is the first zone in the domain, $\rho v^{2}$ represent the ram pressure and $\rho gy$ is the vertical pressure at the guard cells. The others thermodynamic variables are calculated from the equation of state.
For the magnetic field, the two lateral sides are also treated as periodic boundaries, while at the bottom the field is frozen from the initial condition, i.e., the two feet of the loop are 
anchored into the neutron star and no field can be pushed into the star by the accretion. On the top boundary the magnetic field is set to zero, i.e., we assume the accreting matter to be non-magnetized.

On the other hand, our initialization sets the matter in free-fall onto the magnetic loop and we allow the code to find the correct radial profiles of density, pressure, velocity and magnetic field, once the system has been relaxed. Also, we can find a relationship between the shock radius in the cartesian accretion column projection, $y_\mathrm{sh}$, and the accretion rate from analytical considerations. Neutrino losses are dominated mainly by $e^\pm$ pair annihilation for which a simple rate is given by \cite{Dicus1972} as $\dot{\epsilon}_{\nu}\simeq 0.97\times10^{25}(T/\mathrm{MeV})^{9}\:\mathrm{erg\, s^{-1}\, cm^{-3}}$. For given $M$ and $R$, and a fixed $\dot{M}$, the vertical location of the accretion shock is controlled by energy balance between the accretion power and the integrated neutrino losses, per unit neutron star surface area,
\begin{equation}
\frac{GM\dot{M}}{R} = \int_R^\infty \dot{\epsilon}_{\nu}(y) dy.
\label{Eq:Ebalance}
\end{equation}
Due to the resulting strong $y$ dependence of $\dot{\epsilon}_{\nu}$, the value of the upper limit in the integral of Equation~(\ref{Eq:Ebalance}) is not very important and it fixes the height of the accretion shock in the cartesian column as

\begin{equation}
y_\mathrm{sh} \simeq 5.1 \times 10^6 \, (\dot{M}_0/\dot{M})^{10/63} \; \mathrm{cm} \, \simeq 4.2 \times 10^6 \; \mathrm{cm}\,,
\label{eq:ysh}
\end{equation}

\noindent where $\dot{M}_0\sim 340 \; M_\odot \, \mathrm{yr}^{-1}$ is a fiducial accretion rate estimated for SN1987A. The hypercritical accretion rate $\dot{M}$, for the present case can be estimated following \cite{1989ApJ...346..847C} which is written as  
\begin{equation}
\dot{M}\simeq1.2\times10^{-4}\left(\frac{t_{0}}{10\mathrm{\,days}}\right)^{-\frac{1}{3}}\left(\frac{t}{\mathrm{yr}}\right)^{-\frac{5}{3}}\mathrm{\:M_{\odot}\,yr^{-1}}\,.
\end{equation}

\noindent It was obtained considering that because the increasing transparency of the expanding gas, the flow eventually becomes ballistic and the accretion rate is expected to drop as $t^{-5/3}$. As was pointed by \cite{woo93}, the late accretion rate depends on the time $t_{0}$, at which this transition is made and for the parameters of a compact remnant like the Kes 79 case, it can be estimated as $\sim 5-10$ days. It is estimated that the hypercritical phase takes on the order of hours.

In our simulations we use the Adaptive Mesh Refinement capabilities of the FLASH code with six levels of refinement, which implies an effective resolution of $512 \times 10240$ zones in the computational domain. The time step in the FLASH code is adaptive and depends on local conditions. Typically, the time resolution of the simulations is $dt=10^{-6}$ s.

\section{Numerical Results}\label{sec-FLASH}
Our simulations ran for various hundred milliseconds. We found that for the estimated hyperaccretion rate for the Kes 79 scenario, the magnetic field is submerged into the new crust very fast.  Neglecting convection effects, the timescale required for the quasi-stationary solution to set-in is a few sound crossing times, $t_{\rm cross}\simeq r_{\rm shock}/c_{\rm s}$.  For a shock radius of $\simeq 50$~km and $c_{\rm s}\simeq c/10$, it is $t_{\rm cross}\simeq 1-2$~ms. 
The simulations presented here run for almost hundreds of ms, so this is established quite fast.  The timescale for convection is of course much longer,  and will depend on the equilibrium between infall and cooling at the base of the envelope.  In Fig. 1 we show color maps of pressure, density (on a logarithmic scale) and the magnetic field (on a linear scale), for the Kes 79 scenario, when quasi-hydrostatic envelope has been established ($t\simeq 150$ ms).  We found that in the density and pressure contrast the piling up of accreted matter close to the neutron star surface is very notorious (an effect not
accounted for in the analytical approach). In addition, although the system has reached a quasi-stationary state,
little turbulent motion remains in the system, just above the neutron star surface, which is clearer in the magnetic field contrast. It can be explained due to the presence of the confined strong magnetic field and because of the periodic boundary condition on the vertical sides, matter can freely flow in the horizontal direction, as well as bounce off the neutron star surface, which prevents a full stationary state from being reached.
In our simulations we obtain a value of $y_\mathrm{sh} \simeq 4.0 \times 10^6 \; \mathrm{cm}$, in agreement with the analytical approach. It is worth noting that a stationary envelope, in quasi-hydrostatic equilibrium, will expand, or shrink, so that the physical conditions at its base allow neutrinos to carry away all the energy injected by the accretion. Once emitted, neutrinos will act as an energy sink provided the material is optically thin to them. It causes an imbalance in pressure at the base of the column accretion. This fact allows the piling up of matter onto the neutron star surface building a new crust with the bulk of the magnetic field buried by the hyperaccretion in the same height-scale. For the present scenario, the high density due to the strong accretion submerges the magnetic field into the crust deep, quickly and efficiently.
In Fig. 2 we show color maps of ram-pressure (on a logarithmic scale) and the ratios of ram-pressure to magnetic-pressure and magnetic-pressure to thermal-pressure (on a linear scale) for a comparative analysis. Here, the ram-pressure is defined as $p_{ram}=\rho v^2$ and the magnetic-pressure is, basically, $p_{B}=B^2/8\pi$. The thermal pressure is obtained directly from the equation of state.
In the ram-pressure panel, we show the quasi-hydrostatic envelope well established with the in-falling matter raining over it. We note that the flow smooths gradually as the accretion shock stabilizes. Instabilities of the Rayleigh-Taylor type are present in this regime, but they disappear when the system reaches equilibrium.
The initial magnetic loop configuration was immediately torn apart by the reverse shock and the initial strong convection which is dragging the field with it within the developing envelope ($t=1$~ms). As expected, the ram-pressure is dominant inside the envelope, mainly near the stellar surface. The free-fall region has a ram-pressure two orders of magnitude lower. The magnetic-pressure to ram-pressure panel is confirming the dominance of the ram-pressure on the magnetic-pressure in most of the computational domain, except in the region where the magnetic field remain confined, because there these pressures are comparable. Nevertheless, as the matter continues to fall during the whole hypercritical phase, the ram-pressure eventually overtakes the magnetic pressure, submerging the bulk of the magnetic field in the new crust formed with the accreted matter. In \cite{Bernal2013}, the authors performed numerical simulations with similar and larger accretion rates and found that the magnetic field submergence is more notorious and efficient for larger accretion rates.
To compare the magnetic pressure $p_{B} = B^{2}/8\pi \sim10^{23} (B/10^{12} \mathrm{G})^2$ to the thermal-pressure $p$, when the envelope in quasi-hydrostatic equilibrium has been established, we can use its previous definitions to deduce,

\begin{equation}
p = 3.26 \times 10^{28}  (\dot{M}/\dot{M}_0)^{4/9} (10 \,\mathrm{km}/y)^4 \, \mathrm{dyne/cm^2} \, .
\end{equation}
It is worth noting that the Kes 79 hyperaccretion rate is  always in the regime $p_{B} \ll p$. This result has been confirmed by comparing the magnetic-pressure to thermal-pressure panel.  Although in the first stages, the system has a turbulent dynamics, the magnetic field, having its feet frozen into the neutron star surface, always remains confined below the accretion shock. After 40~ms, the magnetic field begins to be submerged and is trapped into the material that is pilling up onto the neutron star surface. The matter outside the accretion shock is still falling with constant accretion rate, but it is the fluid inside the envelope, with its much lower downward velocity, which is nevertheless responsible for the magnetic field submergence. After 150 ms, the submergence is completed, with a maximum strength of the magnetic field $\simeq 6 \times10^{12}$G.\\
In Fig. 3 we show the radial profiles of density, pressure and temperature (on a logarithmic scale), as well as the magnetic field magnitude (on a linear scale), when the quasi-hydrostatic envelope has been established. In the density and pressure profiles the piling up of matter close to the neutron star surface is very notorious, but in the temperature profile this behavior is less notorious. In addition, the magnetic field has been amplified since its seminal value by a factor of six. This may be due to a turbulent dynamo mechanism acting on the crust. However, this hypothesis requires further study that is beyond the scope of this paper. Of course, the question of how the flow approaches the steady state in a time-depending situation remains an open problem. However, numerical simulations with more refinement, more physical ingredients and more degrees of freedom in the system, offer new and unique insight into this problem. As noted by \cite{Bernal2010}, the study of quasi-hydrostatic envelopes around neutron stars poses a challenge. It is as a runaway process in which the neutrino losses lead to the gravitational contraction of the envelope, increasing the pressure at the base of the envelope, which increases the neutrino losses. \\
The neutrino emission is strongly concentrated within the first kilometer above the stellar surface due to the strong temperature dependence of the emissivity. In Figure \ref{fig6} we show a color map of the neutrino emissivity, in linear scale, when the quasi-stationary state is reached. The mean emissivity in such region is $\dot{\epsilon}_{\nu}\simeq 2\times 10^{31}\:\mathrm{erg\, s^{-1}\, cm^{-3}}$ and the mean neutrino luminosity is then $L_{\nu}=\dot{\epsilon}_{\nu}\times V \simeq 8\times 10^{48}\:\mathrm{erg\, s^{-1}}$, with the volume of the region $V=(2\times 10^{6})^{2} \times (10^{5})=4\times 10^{17}$ cm$^{3}$.
\section{Neutrinos}\label{sec-Neutrinos}
\subsection{Events expected in the Hyper-Kamiokande Experiment}

The HK observatory will be the third generation of water Cherenkov detector in Kamioka, designed for a vast variety of neutrino studies.  At a distance of 8 km from its predecessor SK,  the HK detector will be built as the Tochibora mine of the Kamioka Mining and Smelting Company, near Kamioka town in the Gifu Prefecture, Japan.  It will be composed of two separate caverns, each having an egg-shaped cross section of 48 meters wide, 54 meters tall, and 250 meters long. The entire array will be made up of 99.000 PMTs, uniformly surrounding the region. It will have a total (fiducial) mass of 0.99 (0.56) million metric tons, approximately 20 (25) times larger than that of SK.  Among the physical potentials of this detector, SK will be focused on the detection of astrophysical neutrinos and the studies of neutrino oscillation parameters \citep{2011arXiv1109.3262A, 2014arXiv1412.4673H}.\\
It is possible to estimate the numbers of events to be expected in this observatory. The number of events to be expected can be written as \citep{2016arXiv160704633F, 2016arXiv160500571F}


\be
N_{ev}=t\,V N_A\,  \rho_N  \int_{E'} \sigma^{\bar{\nu}_ep}_{cc} \frac{dN}{dE}\,dE
\ee

\noindent where $V$ is the effective water volume, $N_A=6.022\times 10^{23}$ g$^{-1}$ is the Avogadro's number, $\rho_N=2/18\, {\rm g\, cm^{-3}}$ is the nucleon density, $ \sigma^{\bar{\nu}_ep}_{cc}\simeq 9\times 10^{-44}\,E^2_{\bar{\nu}_e}/MeV^2$  is the cross section, $t$ is the observed time \citep{2015APh....70...54F} and $dN/dE$ is the neutrino spectrum.  Regarding the relationship between the  neutrino luminosity $L$ and flux $F$, $L=4\pi d^2_z  F<E>=4\pi d^2_z   E^2 dN/dE$  \citep{2014ApJ...783...44F, 2014MNRAS.441.1209F}  and approximation of the time-integrated average energy  $<E_{\bar{\nu}_e}>$ and time, then this number is 

\bary\label{num_Neu}
N_{ev}&\simeq&\frac{t}{<E_{\bar{\nu}_e}>}V N_A\,  \rho_N  \sigma^{\bar{\nu}_ep}_{cc} <E_{\bar{\nu}_e}>^2\frac{dN}{dE}\cr
&\simeq&\frac{t}{4\pi d^2_z <E_{\bar{\nu}_e}>}V N_A\,  \rho_N  \sigma^{\bar{\nu}_ep}_{cc}\,L_{\bar{\nu}_e.}
\eary
where $d_z$ is the distance from neutrino production to Earth.
\subsection{Neutrino effective potential}
Neutrinos generated during this hypercritical phase and going through the star will oscillate in each zone depending on the neutrino effective potential.
\subsubsection{Zone I}
On the neutron star surface,  the plasma is magnetized $1.0\times10^{12}\,< B<\, 5.0\times 10^{12}$ G (see Fig. \ref{fig3}) and  thermalized at  $1.3\, <T <\, 7$ MeV (Fig. \ref{fig3}).
 In this regime,  the neutrino effective potential  is written as \citep{2014ApJ...787..140F, 2015MNRAS.451..455F,2015APh....71....1F}
\begin{eqnarray}\label{Veffm}
V_{eff,m}=\frac{\sqrt2\,G_F\,m_e^3 B}{\pi^2\,B_c}\biggr[\sum^{\infty}_{l=0}(-1)^l\sinh\alpha_l   \left[F_m-G_m\cos\varphi \right]\nonumber\\
-4\frac{m^2_e}{m^2_W}\,\frac{E_\nu}{m_e}\sum^\infty_{l=0}(-1)^l\cosh\alpha_l  \left[J_m-H_m\cos\varphi \right]  \biggr],
 \end{eqnarray}
where $B_c=4.4\times 10^{13}$ G, $G_F$ is Fermi constant , $m_W$ is the W-boson mass and the functions F$_m$, G$_m$, J$_m$, H$_m$ are 
\bary
F_m&=&\biggl(1+2\frac{E^2_\nu}{m^2_W}\biggr)K_1(\sigma_l)+2\sum^\infty_{n=1}\lambda_n\biggl(1+\frac{E^2_\nu}{m^2_W}  \biggr)K_1(\sigma_l\lambda_n)\nonumber\\
G_m&=&\biggl(1-2\frac{E^2_\nu}{m^2_W}\biggr)K_1(\sigma_l)-2\sum^\infty_{n=1}\lambda_n\frac{E^2_\nu}{m^2_W} K_1(\sigma_l\lambda_n)\nonumber\\
J_m&=& \frac34 K_0(\sigma_l)+\frac{K_1(\sigma_l)}{\sigma_l}+\sum^\infty_{n=1}\lambda^2_n\biggl[K_0(\sigma_l\lambda)+\frac{K_1(\sigma_l\lambda)}{\sigma_l\lambda}\nonumber\\
&&\hspace{5.2cm}-\frac{K_0(\sigma_l\lambda)}{2\lambda^2_n}  \biggr]\nonumber\\
H_m&=& \frac{K_1(\sigma_l)}{\sigma_l}+ \sum^\infty_{n=1}\lambda^2_n   \biggl[\frac{K_1(\sigma_l\lambda)}{\sigma_l\lambda} - \frac{K_0(\sigma_l\lambda)}{2\lambda^2_n}  \biggr]
\eary

where $\lambda^2_n=1+2\,n\,B/B_c $, K$_i$ is the modified Bessel function of integral order i, $\alpha_l=\beta\mu(l+1)$ and $\sigma_l=\beta m_e(l+1)$.  It is important to clarify that as the magnetic field decreases the effective potential will depend less on the Landau levels.
\subsubsection{Zone II, III and IV}
In zones II, III and IV, neutrinos will undergo different effective potentials. It can be written as
\be
V_{eff}=\sqrt2 G_F\, N_A\,\rho(r)\,Y_e
\ee
where $Y_e$ is the number of electron per nucleon and $\rho(r)$ is given in table \ref{table1}.
%
\subsection{Neutrino Oscillation}

Measurements of the fluxes at solar, atmospheric and accelerator neutrinos have showed overwhelming evidence of neutrino oscillations and then neutrino masses and mixing. To make a full analysis,  we are going to show the important quantities to involve in neutrino oscillations in vacuum and matter as well as the two and three- mixing parameters. The two-mixing parameters are related as follows:\\
Solar Experiments: A two-flavor neutrino oscillation analysis yielded $\delta m^2=(5.6^{+1.9}_{-1.4})\times 10^{-5}\,{\rm eV^2}$ and $\tan^2\theta=0.427^{+0.033}_{-0.029}$\citep{aha11}.\\
Atmospheric Experiments: Under a two-flavor disappearance model with separate mixing parameters between neutrinos and antineutrinos the following parameters for the SK-I + II + III data $\delta m^2=(2.1^{+0.9}_{-0.4})\times 10^{-3}\,{\rm eV^2}$ and $\sin^22\theta=1.0^{+0.00}_{-0.07}$ were found \citep{abe11a}.\\
Accelerator Experiments: \cite{chu02} found two well defined regions of oscillation parameters with either $\delta m^2  \approx  7\, {\rm eV^2}$ or $\delta m^2 < 1\, {\rm eV^2} $ compatible with both LAND and KARMEN experiments, for the complementary confidence and the angle mixing is $\sin^2\theta=0.0049$. In addition, MiniBooNE found evidence of oscillations in the 0.1 to 1.0 eV$^2$, which are consistent with LSND results \citep{1998PhRvL..81.1774, 1996PhRvL..77.3082A}.   \\
The combining solar, atmospheric and accelerator parameters are related in three-mixing parameters as follows \citep{aha11,wen10}:
\bary\label{3parosc}
{\rm for}&&\,\,\sin^2_{13} < 0.053: \delta m_{21}^2= (7.41^{+0.21}_{-0.19})\times 10^{-5}\,{\rm eV^2}\, {\rm and}\cr
&&\hspace{3.7cm} \tan^2\theta_{12}=0.446^{+0.030}_{-0.029}\cr
{\rm for}&&\,\,\sin^2_{13} < 0.04: \delta m_{23}^2=(2.1^{+0.5}_{-0.2})\times 10^{-3}\,{\rm eV^2}{\rm and}\cr
&& \hspace{3.8cm} \sin^2\theta_{23}=0.50^{+0.083}_{-0.093}
\eary

\subsubsection{In Vacuum}

Neutrino oscillation in vacuum would arise if neutrino were massive and mixed. For massive neutrinos, the weak eigenstates $\nu_\alpha$ are linear combinations of mass eigenstates {\small $\mid \nu_\alpha> \sum^n_{i=1} U^*_{\alpha i} \mid \nu_i>$}.  After traveling a distance $L\simeq ct$, a neutrino produced with flavor $\alpha$ evolve to {\small $\mid \nu_\alpha (t)> \sum^n_{i=1} U^*_{\alpha i} \mid \nu_i (t)>$}. Then, the transition probability  from a flavor estate $\alpha$ to a flavor state $\beta$ can be written as ${\small P_{\nu_\alpha\to\nu_\beta} =\delta_{\alpha\beta}-4 \sum_{j>i}\,U_{\alpha i}U_{\beta i}U_{\alpha j}U_{\beta i}\,\sin^2 [\delta m^2_{ij} L/(4\, E_\nu )]}$ with $L$ is the distance traveled by the neutrino in reaching Earth (detector).  Using the set of parameters give in eq. (\ref{3parosc}) and averaging the sin term in the probability to $\sim 0.5$ for larger distances L (longer than the solar system) \citep{lea95}, the probability matrix for a neutrino flavor vector of ($\nu_e$, $\nu_\mu$, $\nu_\tau$)$_{source}$ changing to a flavor vector  ($\nu_e$, $\nu_\mu$, $\nu_\tau$)$_{Earth}$ is given as
{\small
\be
{\pmatrix
{
\nu_e   \cr
\nu_\mu   \cr
\nu_\tau   \cr
}_{E}}
=
{\pmatrix
{
0.534143	  & 0.265544	  & 0.200313\cr
 0.265544	  & 0.366436	  &  0.368020\cr
 0.200313	  & 0.368020	  & 0.431667\cr
}}
{\pmatrix
{
\nu_e   \cr
\nu_\mu   \cr
\nu_\tau   \cr
}_{S}}
\label{matrixosc}
\ee
}
where $E$ and $S$, are Earth and source, respectively. 
\subsubsection{In Matter}
\paragraph{Two-Neutrino Mixing.}   The evolution equation for the propagation of neutrinos in the above medium is given by \citep{2016JHEAp...9...25F}
\be
i
{\pmatrix {\dot{\nu}_{e} \cr \dot{\nu}_{\mu}\cr}}
={\pmatrix
{V_{eff,s/m}-\Delta \cos 2\theta & \frac{\Delta}{2}\sin 2\theta \cr
\frac{\Delta}{2}\sin 2\theta  & 0\cr}}
{\pmatrix
{\nu_{e} \cr \nu_{\mu}\cr}},
\ee
where $\Delta=\delta m^2/2E_{\nu}$, $V_{eff,s/m}$ is the effective potential given by eq. (\ref{Veffm}),    $E_{\nu}$ is the neutrino energy and $\theta$ is the neutrino mixing angle.  Here we have considered the neutrino oscillation process $\nu_e\leftrightarrow \nu_{\mu, \tau}$. The transition probability in matter is {\small $P_{\nu_e\rightarrow {\nu_{\mu}{(\nu_\tau)}}}(t) = \frac{\Delta^2 \sin^2 2\theta}{\omega^2}\sin^2\left (\frac{\omega t}{2}\right)$} with $\omega=\sqrt{(V_{eff}-\Delta \cos 2\theta)^2+\Delta^2 \sin^2 2\theta}$.  This probability has an oscillatory behavior, with oscillation length given by
\be
L_{osc}=\frac{L_v}{\sqrt{\cos^2 2\theta (1-\frac{V_{eff}}{\Delta \cos 2\theta}
    )^2+\sin^2 2\theta}},
\label{osclength}
\ee
where $L_v=2\pi/\Delta$ is the vacuum oscillation length.  Satisfying the resonance condition
\be
V_{eff} -  \frac{\delta m^2}{2E_{\nu}} \cos 2\theta = 0,
\label{reso}
\ee
the resonance length can be written as {\small $L_{res}=\frac{L_v}{\sin 2\theta}$}.
\paragraph{Three-Neutrino Mixing.} The neutrino dynamics is determined by the evolution equation in a three-flavor framework which can be written as
\be
i\frac{d\vec{\nu}}{dt}=H\vec{\nu},
\ee
and the state vector in the flavor basis is defined as
\be
\vec{\nu}\equiv(\nu_e,\nu_\mu,\nu_\tau)^T.
\ee
The effective Hamiltonian is \citep{2013MNRAS.tmp.2798F}
\be
H=U\cdot H^d_0\cdot U^\dagger+diag(V_{eff},0,0),
\ee
with
\be
H^d_0=\frac{1}{2E_\nu}diag(-\delta m^2_{21},0,\delta m^2_{32}).
\ee
Here $U$ is the three neutrino mixing matrix given by \citep{gon03,akh04,gon08,gon11}.  The  oscillation length of the transition probability  is given by
\be
l_{osc}=\frac{l_v}{\sqrt{\cos^2 2\theta_{13} (1-\frac{2 E_{\nu} V_e}{\delta m^2_{32} \cos 2\theta_{13}}
    )^2+\sin^2 2\theta_{13}}},
\label{osclength}
\ee
where $l_v=4\pi E_{\nu}/\delta m^2_{32}$ is the vacuum oscillation length. The resonance condition and resonance length are,
\be\label{reso3}
V_{eff}-5\times 10^{-7}\frac{\delta m^2_{32,eV}}{E_{\nu,MeV}}\,\cos2\theta_{13}=0
\ee
and  {\small $l_{res}=\frac{l_v}{\sin 2\theta_{13}}$}.
Considering the adiabatic condition at the resonance, it is expressed as 
{\small
\be
\kappa_{res}\equiv  \frac{2}{\pi}
\left ( \frac{\delta m^2_{32}}{2 E_\nu} \sin 2\theta_{13}\right )^2
\left (\frac{dV_{eff}}{dr}\right)^{-1} \ge 1
\label{adbcon}
\ee
}
\section{Conclusions}\label{sec-Results}
In the present work we have studied the dynamics of hyperaccretion (with neutrino emission) onto a newly born neutron star as is the case of Kes 79 scenario. To do this, we have performed numerical 2D-MHD simulations using the AMR FLASH method. The code allows to capture the rich morphology and the main characteristics of the fundamental physical processes in detail. We included a custom routine in the code that take into account several neutrino cooling processes, which are active depending of strict criteria of temperature and density of the model in the computational domain. With this, we found that for the estimated hyperaccretion rate for the Kes 79 scenario, the bulk of magnetic field is submerged efficiently into the new crust formed by the accreted matter during this regime. Additionally, we found that the code reproduce the radial profile of the main thermodynamical parameters of the hypercritical regime, including the accretion shock radius, which it is in good agreement with the value analytically estimated. It is noteworthy that a (completely relaxed) steady state could not be reached, within the parameters of the simulation, due to various factors: periodicity of the lateral boundaries, effects of buoyancy and magnetic stresses near the stellar surface, among others. Also, although most of the magnetic field is confined/submerged in the stellar crust, residual weak magnetic field is present within the quasi-hydrostatic envelope but that does not influence the dynamics of the system.

Once the hypercritical phase ends, the story for the magnetic field continues. \cite{1995ApJ...440L..77M} found that after hyperaccretion stopped, the bulk of the submerged magnetic field could diffuse back to the surface by ohmic processes and depending on the amount of accreted matter, the submergence could be so deep that the neutron star may appear and remain unmagnetized for centuries. It is possible that, in addition to the ohmic diffusion of the bulk of the magnetic field, some portion of the confined magnetic field may be pushed into the neutron star by the hyperaccretion, and it may be crystallized in the mantle. The presence of a crystal lattice of atomic nuclei in the crust is mandatory for modeling of the subsequent radio-pulsar glitches (see \cite{1999A26A...345..847G} and references therein). Presence of solid crust enables excitation of toroidal modes of oscillations. The toroidal modes in a completely fluid star have all zero frequency, but the presence of a solid crust gives them nonzero frequencies in the range of kHz.
On the other hand, requiring the value of distance $d_z=7.$1 kpc \citep{1998ApJ...504..761C}, neutrino Luminosity  $L_{\bar{\nu}_e}=(8.4\pm 0.4)\times10^{48}\:\mathrm{erg\, s^{-1}}$ (this work, Fig. \ref{fig6}), effective volume of HK, $V\simeq 0.56\times 10^{12}\,{\rm cm^3}$ \citep{2014arXiv1412.4673H}  and the average neutrino energy  $<E_{\bar{\nu}_e}>\simeq 7\, {\rm  MeV}$, from eq. (\ref{num_Neu}) we obtain that the number of events that could have been expected from the hypercritical phase  on Hyper-Kamiokande is 733$\pm$364.   In addition,  we compute the number of initial neutrino burst expected during the neutron star formation.  Taking into account a temperature $T \approx 4\pm1$ MeV \citep{Giunti-Kim-2007},  a duration of the neutrino pulse in the hypercritical phase $t\simeq10^{3}$ s, the average neutrino energy  $<E_{\bar{\nu}_e}>\simeq  13.5\pm 3.2\, {\rm  MeV}$ \citep{Giunti-Kim-2007} and a total fluence equivalent of Kes 79 is $\Phi\approx(1.25\pm 0.52) \times 10^{11}\, \bar{\nu}_e \,{\rm cm^{-2}}$ \citep{2004mnpa.book.....M, 1989neas.book.....B}, then the total number of neutrinos emitted from Kes 79 would be $N_{tot}=6\,\Phi\, 4\pi\, d_z^2\approx(8.97\pm 3.58) \times 10^{57}$.  Similarly, we can compute the total radiated luminosity  corresponding to the binding energy of the neutron star $L_\nu\approx \frac{N_{tot}}{t}\times <E_{\bar{\nu}_e}>\approx (2.16 \pm0.87) \times 10^{52}$ erg/s.  Regarding the effective volume $V\simeq 0.56\times 10^{12}\,{\rm cm^3}$ \citep{2014arXiv1412.4673H}, from eq. (\ref{num_Neu}) we get  $N_{ev}\simeq 1129\pm 475$ events expected during the neutron star formation in a neutrino detector as HK experiment. Comparing  the number of neutrinos expected during the neutron star formation and the hyperaccretion phase, we obtain 1.5 events.\\
Neutrinos generated at the hypercritical accretion phase will oscillate in their ways due to electron density in zones I, II, III and IV, and after in vacuum into Earth. In zone I, the thermal plasma is endowed with a magnetic field $\sim$ (1 - 6)$\times 10^{12}$ G and thermalized $\sim$ (1 - 8) MeV. Taking into account the neutrino effective potential given in zone I, we plot the neutrino effective potential as a function of magnetic field, angle, temperature and chemical potential.  From fig. \ref{fig4} can be observed  the positivity of the effective potential ($V_{eff}>$ 0), therefore neutrinos can oscillate resonantly. Taking into account the two and three- neutrino mixing parameters we analyze the resonance condition, as shown in Figure \ref{fig5}.  Recently,  \citet{2014MNRAS.442..239F} showed that neutrinos can oscillate resonantly due to  the density profiles of the collapsing material surrounding the progenitor. In zone II, III and IV,  the author showed that neutrinos can oscillate resonantly and computed the survival and conversion probabilities for the active-active ($\nu_{e,\mu,\tau} \leftrightarrow \nu_{e,\mu,\tau}$) neutrino oscillations in each region. Taking into account the oscillation probabilities in each region and in the vacuum (on its path to Earth), we calculate the flavor ratio expected on Earth for neutrino energies of $E_{\nu}=1$ MeV, 1 MeV, 3 MeV, 5 MeV and 7 MeV, as shown in table \ref{Table}. In this table we can see a small deviation from the standard ratio flavor 1:1:1.   In this calculation we take into account that for neutrino cooling processes (electron-positron annihilation, inverse beta decay, nucleonic bremsstrahlung and plasmons), only  inverse beta decay is the one producing electron neutrino. It is worth noting that our calculations of resonant oscillations were performed for neutrinos instead of anti-neutrinos, due to the positivity of the neutrino effective potential.\\

As a final remark, we can say that the present numerical studies of these phenomena are necessary and very important and its results can give us a glimpse of the complex dynamics around the newly born neutron stars, moments after the core-collapse supernova explosion. Although we use the Kes 79 as a particular case of the hidden magnetic field scenario, this method can be applied to the various CCOs scenarios. Nevertheless, the eventual growing of the bulk magnetic field post-hyperaccretion, the possible magnetic field amplification by turbulent dynamo and the magnetic field crystallization processes inside neutron star require more detailed studies that are outside the scope of this paper.
\begin{table*}      
\begin{center}
\caption[]{The neutrino flavor ratio in each region of the hypercritical phase for  $E_{\nu}=$ 1, 3, 5 and 7 MeV.}\label{Table}
\begin{minipage}{126mm}
\begin{tabular}{lccccccccc}
\hline
 
 $E_{\nu}$  & On the NS surface & Accretion material & Free fall zone& Outer layers & On Earth \\\hline \hline 

{\small 1}     &  {\small 1.2:0.9:0.9}  &  {\small 1.186:0.907:	0.907} & {\small 1.152:0.924:0.924} &  {\small 1.120:0.940:0.940}  &  {\small 1.036:0.988:0.976} \\\hline

{\small 3}   & {\small 1.2:0.9:0.9}  &  {\small 1.161:0.919:0.919}&  {\small 1.130:0.935:0.935} &  {\small 1.132:0.934:	0.934}  &  {\small 1.039:0.987:0.974}\\\hline

{\small 5}  & {\small 1.2:0.9:0.9} &  {\small 1.159:0.921:0.921} &  {\small 1.127:0.937:0.937} &  {\small 1.123:0.938:0.938}  &  {\small 1.037:0.987:0.975}   \\\hline

{\small 7}  &  {\small 1.2:0.9:0.9}  &   {\small 1.172: 0.914:0.914}  &   {\small 1.142:0.929:0.929} &  {\small 1.135:0.933:0.933}  &  {\small 1.041:0.987:0.973}  \\\hline
 
\end{tabular}
\end{minipage}
\end{center}
\end{table*}

\section*{Acknowledgements}
We thank the anonymous referee for a critical reading of the paper and valuable suggestions that helped improve the quality and clarity of this work. We also thank to Dany Page, John Beacom and W. H. Lee for useful discussions.
C.B. acknowledges to CAPES--Brazil by the postdoctoral fellowship received through the Science Without Borders program. This work was supported by the projects IG100414 and Conacyt 101958. The software used in this work was in part developed by the DOE NNSA-ASC OASCR FLASH Center at the University of Chicago. \\
%
%

%
%
\clearpage
\begin{figure*}
\centering
\includegraphics[width=0.84\textwidth]{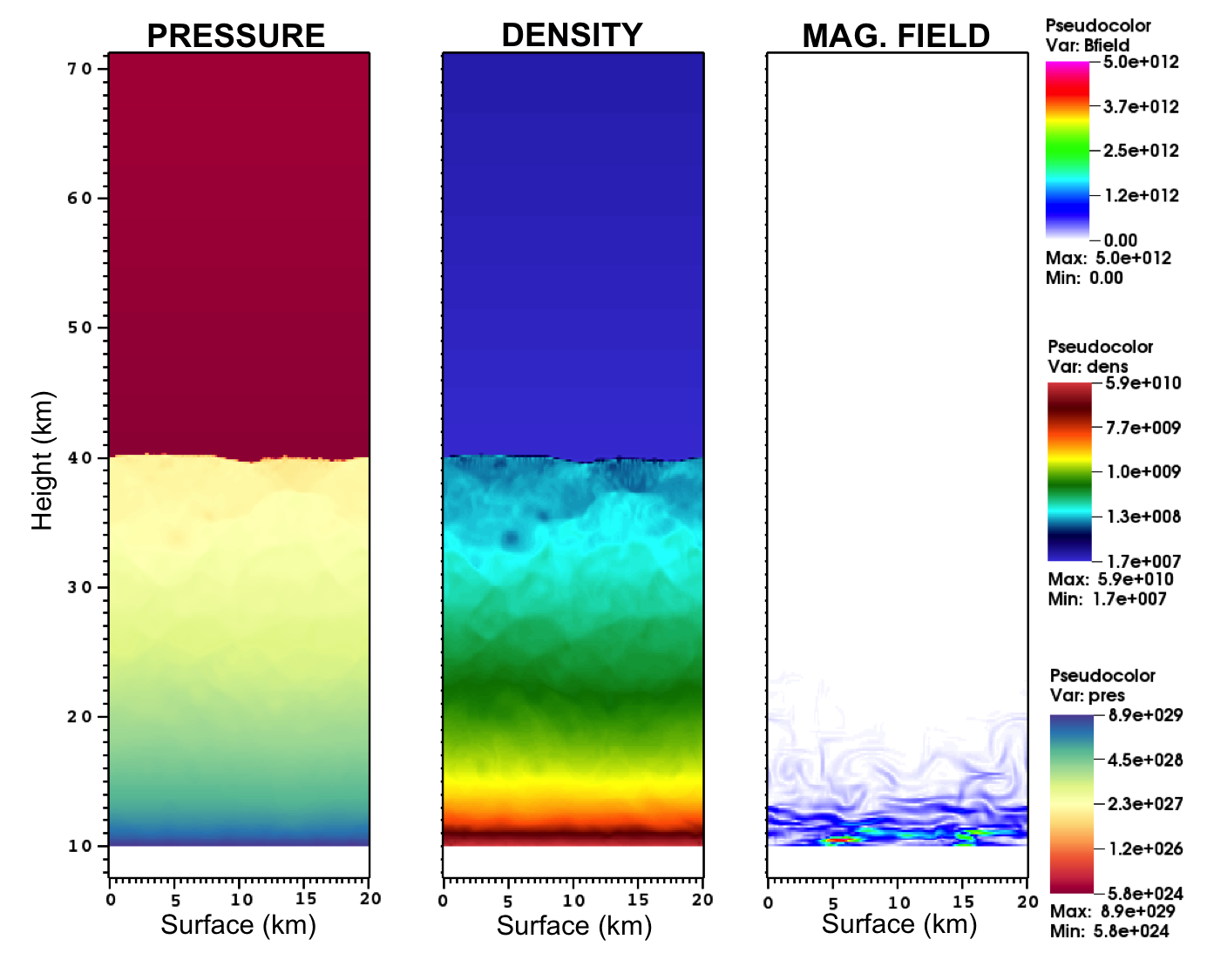}
\caption{Color maps of pressure and density (on a logarithmic scale), as well as the magnetic field magnitude (on a linear scale) for the Kes 79 hyperaccretion rate. We show 2D MHD simulations,  when a quasi-hydrostatic envelope has been established ($t=150$ ms). Note the formation of the new crust and the submergence of the magnetic field in the same highscale.}
\label{fig1}
\end{figure*}
\begin{figure*}
\centering
\includegraphics[width=0.84\textwidth]{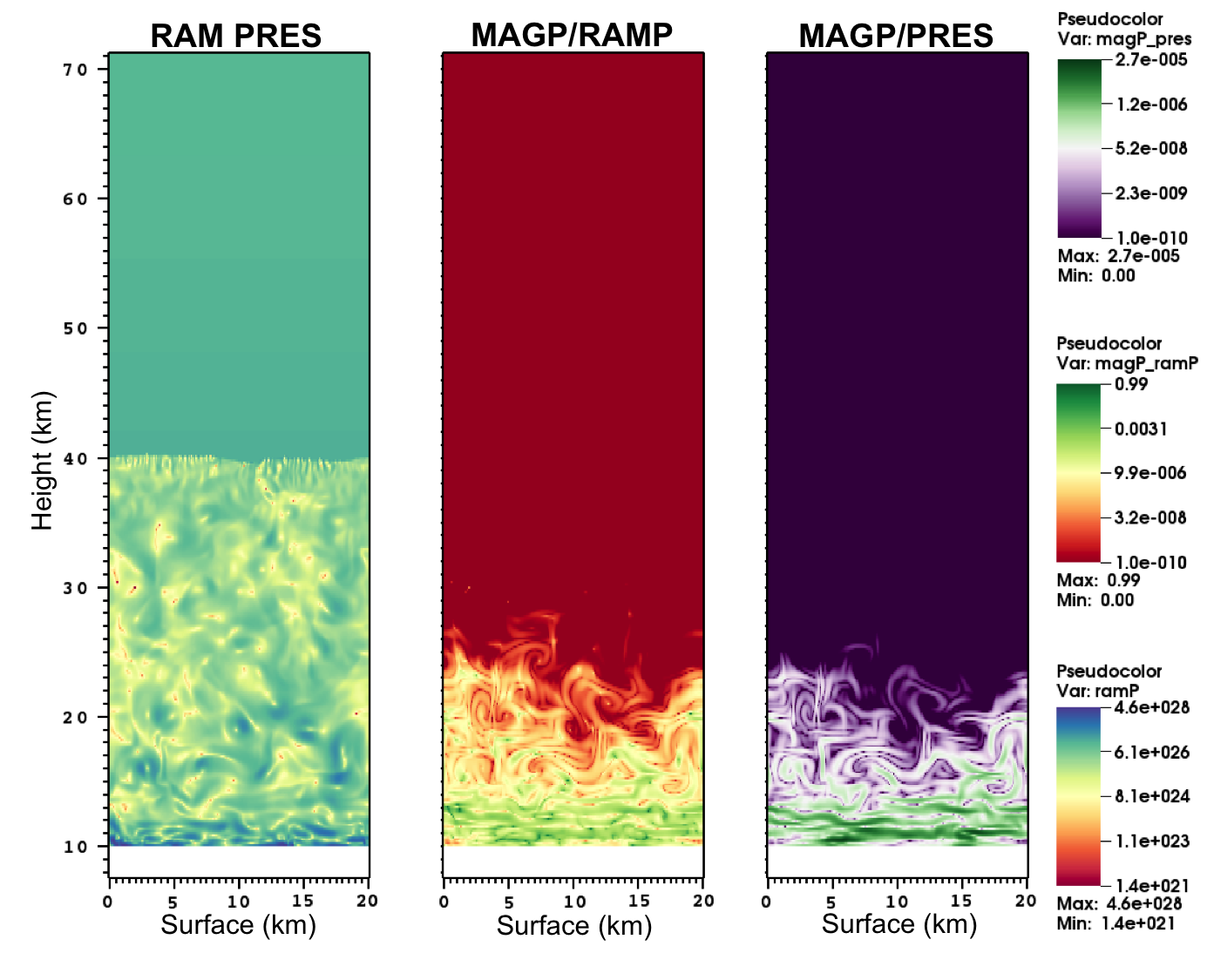}
\caption{Color maps of ram-pressure, ratios of magnetic-pressure to ram-pressure and magnetic-pressure to thermal pressure, for the Kes 79 hyperaccretion rate for a comparative analysis, in $t=150$ ms. Note the dominance of the ram-pressure over the magnetic pressure allowing the magnetic field submergence.}
\label{fig2}
\end{figure*}
\begin{figure*}
\centering
\includegraphics[width=0.84\textwidth]{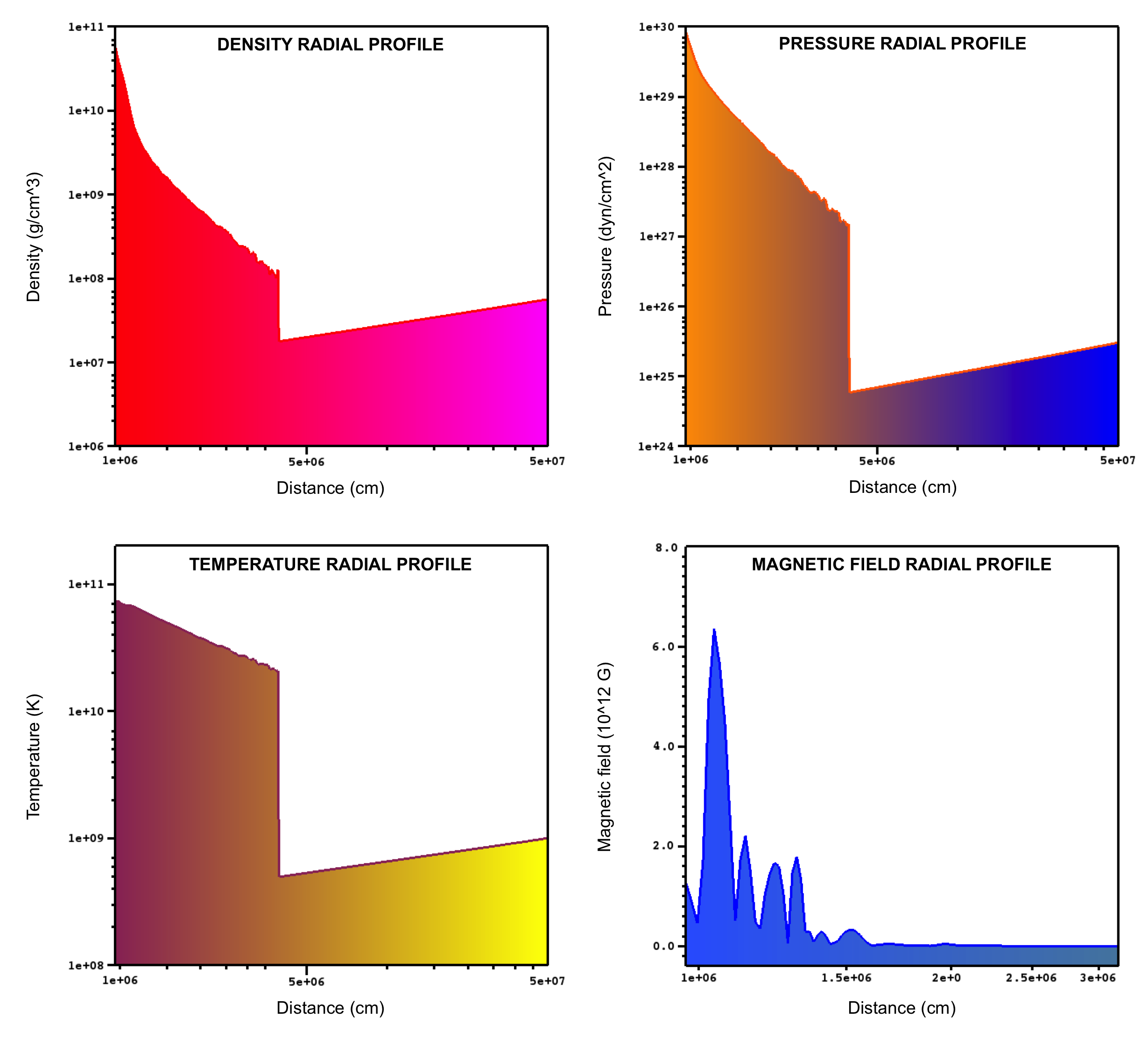}
\caption{Radial profiles of density, pressure, temperature (on a logarithmic scale), and magnetic field (on a linear scale) for the Kes 79 hyperaccretion rate when a quasi-hydrostatic envelope has been established ($t=150$ ms). The free-fall region, the quasi-hydrostatic envelope and the new crust are present.}
\label{fig3}
\end{figure*}
\begin{figure*}
\centering
\includegraphics[width=0.54\textwidth]{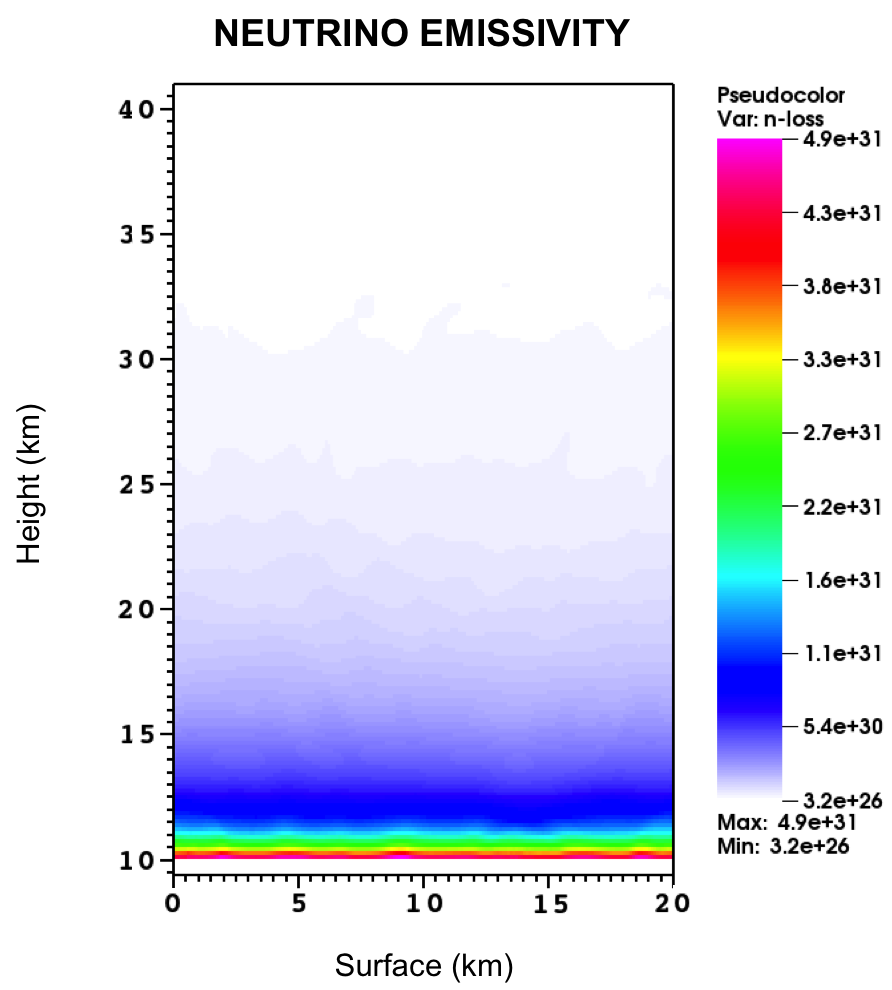}
\caption{Color map, in linear scale, of the neutrino emissivity when the quasi-stationary state is reached $(t=150)$ ms. Note the strong emissivity concentred very close of the stellar surface.}
\label{fig6}
\end{figure*}
\begin{figure*}
\centering
\includegraphics[width=\textwidth]{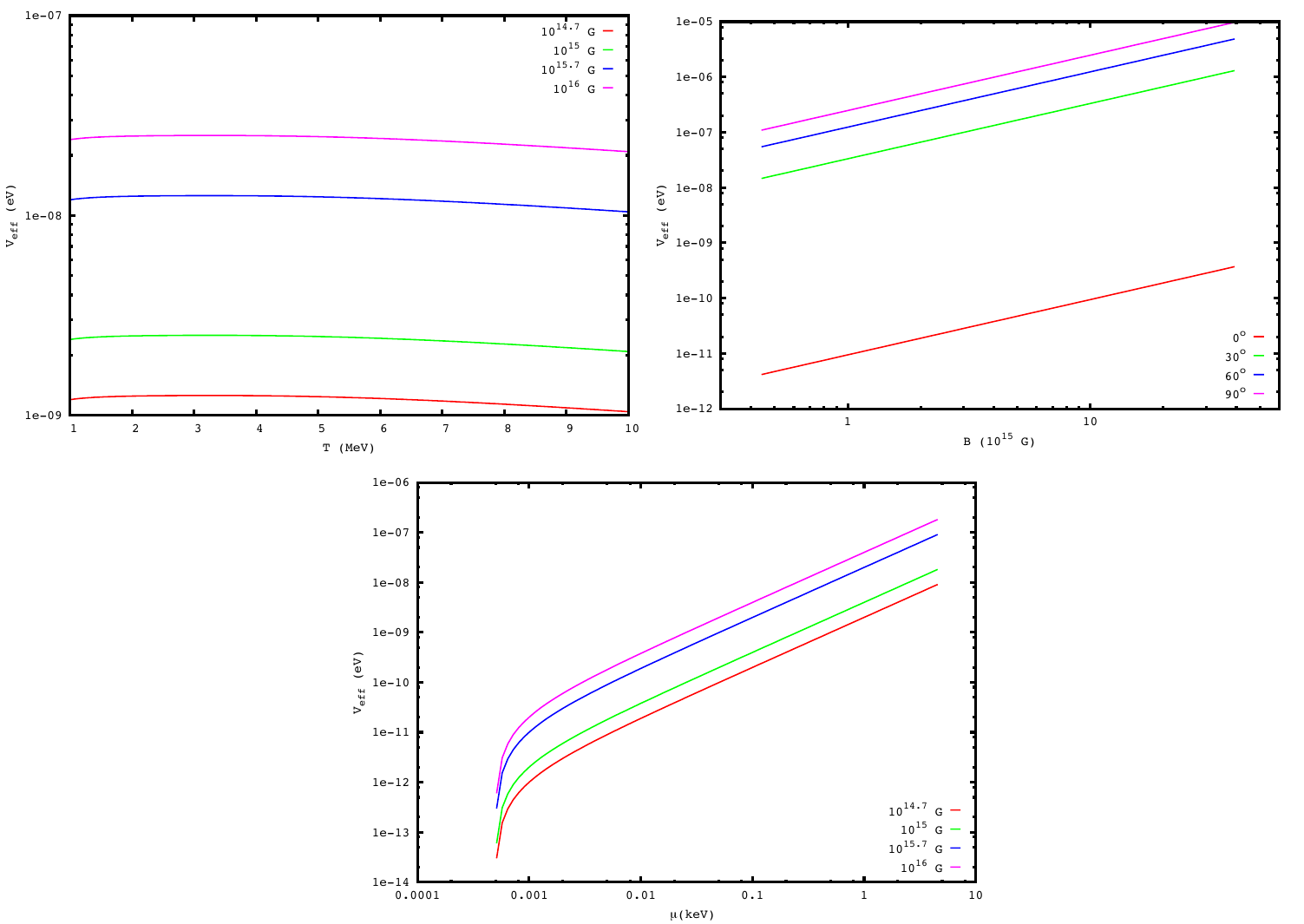}
\caption{Neutrino effective potential in the moderate magnetic field regime as a function of temperature (top left), magnetic field (top right) and chemical potential (bottom).}
\label{fig4}
\end{figure*}
\begin{figure*}
\centering
\includegraphics[width=0.84\textwidth]{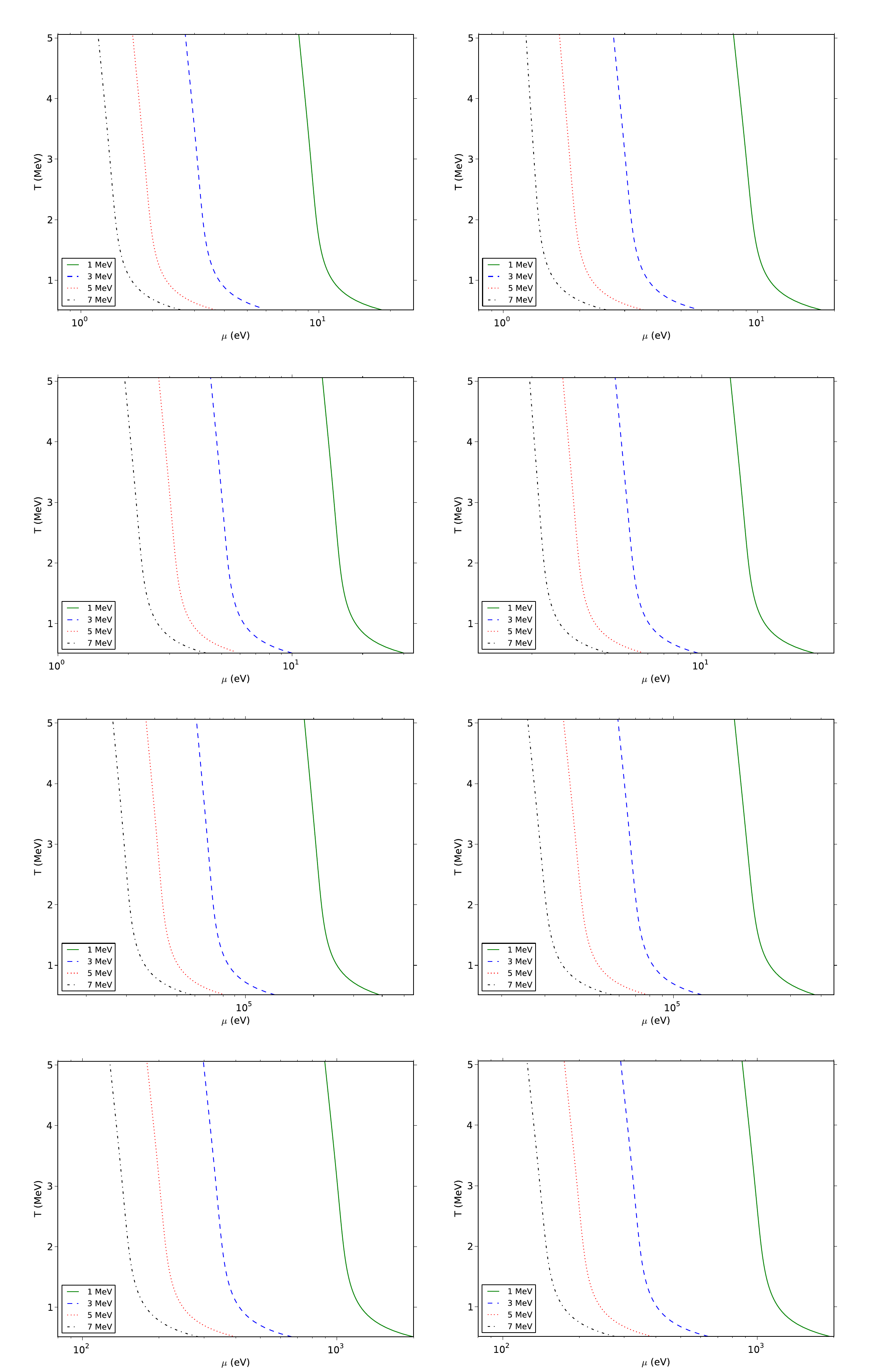}
\caption{Contour plots of temperature (T) and chemical potential  ($\mu$) as a function of neutrino energy for which the resonance condition is satisfied. We have applied the neutrino effective potential at the moderate field limit, two values of angles $\varphi=0^\circ$ (left column) and $\varphi=90^\circ$ (right column).   We have used the best fit values of neutrino mixing.  From top to bottom: solar,  atmospheric, accelerator and three flavors.}
\label{fig5}
\end{figure*}
\end{document}